\documentclass[%
 aip,
 amsmath,amssymb,
 reprint,%
]{revtex4-1}

\usepackage{graphicx}
\usepackage{dcolumn}
\usepackage{bm}
\usepackage[utf8]{inputenc}
\usepackage[hidelinks,breaklinks,colorlinks=true,linkcolor=blue,citecolor=blue,urlcolor=blue]{hyperref}
\usepackage[T1]{fontenc}
\usepackage{mathptmx}

\usepackage{etoolbox}
\makeatletter
\let\old@makecaption=\@makecaption
\usepackage{subcaption}
\let\@makecaption=\old@makecaption
\makeatother
\makeatletter
\def\@email#1#2{%
 \endgroup
 \patchcmd{\titleblock@produce}
  {\frontmatter@RRAPformat}
  {\frontmatter@RRAPformat{\produce@RRAP{*#1\href{mailto:#2}{#2}}}\frontmatter@RRAPformat}
  {}{}
}%
\makeatother
\begin{document}

\preprint{AIP/123-QED}

\author{Nikhil R. Agrawal}
\affiliation{ 
Department of Chemical and Biomolecular Engineering, University of California, Berkeley, California 94720, USA
}%
\author{Ravtej Kaur}
\affiliation{ 
Department of Chemical and Biomolecular Engineering, University of California, Berkeley, California 94720, USA
}%
\author{Carlo Carraro}
\affiliation{ 
Department of Chemical and Biomolecular Engineering, University of California, Berkeley, California 94720, USA
}
\author{Rui Wang}
\email{ruiwang325@berkeley.edu}
\affiliation{ 
Department of Chemical and Biomolecular Engineering, University of California, Berkeley, California 94720, USA
}
\affiliation{ 
Materials Sciences Division, Lawrence Berkeley National Laboratory, Berkeley, California 94720, USA
}

\title{Ion correlation driven like-charge attraction in multivalent salt solutions}

\date{\today}

\begin{abstract}
 The electrostatic double layer force is key to determining the stability and self-assembly of charged colloids and many other soft matter systems. Fully understanding the attractive force between two like-charged surfaces remains a great challenge. Here we apply the modified Gaussian renormalized fluctuation theory to study ion correlation-driven like-charge attraction in multivalent salt solutions. The effects of spatially varying ion correlations on the structure of overlapping double layers and their free energy are self-consistently accounted for. In the presence of multivalent salts, increasing surface charge or counterion valency leads to a short-range attraction. We demonstrate that although both overcharging and like-charge attraction are outcomes of ion correlation, there is no causal relationship between them. Our theory also captures the non-monotonic dependence of like-charge attraction on multivalent salt concentration. The reduction of attraction at high salt concentrations could be a contributing factor towards the reentrant stability of charged colloidal suspensions. Our theoretical predictions are consistent with the observations reported in experiments and simulations.
\end{abstract}

\maketitle

\section{\label{sec:level1}Introduction\protect\\}

Interfacial forces determine a wide range of structural and dynamic properties in physical chemistry, colloidal science, soft matter, and biophysical systems\cite{israelachvili2011intermolecular}. Understanding these forces is crucial to explain protein stability\cite{Caccamo_2000,Butler2003IonAttraction, Zhang2008ReentrantCounterions,Nassar2021TheTheory}, self-assembly of colloidal particles\cite{Larsen1997Like-chargeCrystallites,Dinsmore1998Self-assemblyCrystals}, fusion of lipid bilayers\cite{Itskovich1977ElectricCharacteristic, Cevc1990MembraneElectrostatics,Wennerstrom1991IonicCounterions,Kozlov2014MechanismsMembranes,Mukhina2019AttractiveGeometry}, DNA complexes for gene delivery\cite{Gelbart2007DNAInspiredElectrostatics,Solis2007FlexibleCounterattract},  nanofluidics\cite{Schoch2008TransportNanofluidics}, and numerous other phenomena \cite{kornyshevhelices,Besteman2007ChargeIons, Tata2001ColloidalColloids,Levin2002ElectrostaticBiology, Kjellander1990ASolutions}.\par

The starting point for describing the force between two surfaces immersed in an electrolyte solution is the classical Derjaguin-Landau-Verwey-Overbeek (DLVO) theory, which accounts for two contributions: the short-range Van der Waals force and the long-range electrostatic force. This electrostatic force, originating from overlapping electrical double layers (EDLs)\cite{hansenlowenforces}, is described in the DLVO framework by the mean-field Poisson-Boltzmann (PB) theory. For two like-charged surfaces, PB predicts a universally repulsive force. However, numerous experimental \cite{Kekicheff1993ChargeElectrolyte,Zohar2006ShortSurfaces,Kumar2017InteractionsIons} and simulation studies\cite{linselca1999,Linse2000,Wu1999MonteSalts,Angelescu2003MonteAdded,Zhang2016PotentialEffect} have reported that this force becomes attractive at short distances in the presence of multivalent salts. This counterintuitive phenomenon is known as ``like-charge attraction" and cannot be captured even qualitatively by mean-field PB.\par

One crucial factor missing in the Poisson-Boltzmann equation (PB) is the electrostatic correlation between ions. This effect becomes particularly significant in the case of multivalent salts, where the relative strength of Coulombic forces versus thermal forces is amplified. Molecular simulations using only hard-sphere and coulombic force fields have identified the ion correlation effect as the origin of like-charge attraction in multivalent salts\cite{Linse2000,Wu1998InteractionSolutions,Wu1999MonteSalts,lobaskinqamhieh}. Such simulations also observed continuous transition from repulsion to attraction as surface charge increases\cite{Guldbrand1984ElectricalStudy,Pellenq1997ElectrostaticSimulation,IseLikeInteractionw}. The dependence of the strength of attraction on multivalent salt concentration is rather non-trivial and shows non-monotonic behavior. Wu et al.\cite{Wu1999MonteSalts} and Angelescu et al.\cite{Angelescu2003MonteAdded} showed that attraction is most pronounced at an intermediate salt concentration and becomes weaker at high salt concentrations. A related phenomenon observed in experiments is the reentrant stability of charged colloids where the aggregates redissolve at high multivalent salt concentrations\cite{Asor2017CrystallizationNanoparticles, Kumar2017InteractionsIons, Pelta1996DNACobalthexamine}. Furthermore, it is interesting to note that adding monovalent salt to multivalent salts significantly increases the solubility of colloids at low multivalent salt concentrations. However, this effect is negligible when the concentration of multivalent salt is high\cite{Pelta1996DNACobalthexamine, Raspaud1998PrecipitationBehaviour, OlveraDeLaCruz1995PrecipitationSalts, Asor2017CrystallizationNanoparticles}.\par

Over the years many theories have been proposed to understand ion correlation-driven like-charge attraction\cite{Patey1980TheApproximation,Kjellander1988Double-LayerSwelling,Kjellander1988SurfaceElectrolytes,Rouzina1996MacroionCloud, Chu1996AttractiveParticles,Ha1997Counterion-MediatedRods,Arenzon1999SimplePolyions,Diehl1999Density-functionalPlates, Nguyen2000ReentrantCounterions,Grosberg2002Colloquium:Systems,Kudlay2003PrecipitationSolutions,Netz2000BeyondFunctions,Netz2001ElectrostatisticsTheory,Naji2004AttractionLimit, Buyukdagli2017Like-chargeMolecules,Suematsu2018CiteAs, Misra2019TheoryElectrolytes, Chen2020MultivalentPolyelectrolytes,vsamaj2011counter, lobaskinpre2001,gupta_prl_overcharging, gupta_force, wucharging2022}. 
The Strongly Correlated Liquid (SCL) theory assumes a Wigner-crystal lattice of correlated counterions at the surface in contact with a diffuse layer described by mean-field PB\cite{Rouzina1996MacroionCloud,Shklovskii1999ScreeningCharge, Nguyen2000ReentrantCounterions}. Although SCL predicts an attractive force, the model is only valid in the so-called strong coupling limit\cite{SamajCounterionsAttraction,Chen2020MultivalentPolyelectrolytes}. Like-charge attraction is correlated to the overcharging of the EDL\cite{Nguyen2000ReentrantCounterions,Grosberg2002Colloquium:Systems}, where overcharging is defined as the excess accumulation of counterions in the double layer that could lead to a change in the sign of the curvature of the electrostatic potential\cite{Agrawal2022OnLayers,agrawal_aiche}. The subsequent reentrant condensation at high multivalent salt concentration is attributed to the significant overcharging. As per SCL theory, like-charge attraction is most prominent when the surfaces are neutralized, whereas the net electrostatic force could become repulsive when there is either significant undercharging or overcharging. To get repulsion at high salt concentrations, the surfaces have to be significantly overcharged \cite{Nguyen2000ReentrantCounterions}. 
However, simulations by Pai-Yi Hsiao\cite{Hsiao2008OverchargingStudy} showed that the redissolution of polyelectrolyte condensates at high salt concentrations can occur without any inversion of their net charge. In addition, the same simulations also showed that like-charge attraction can happen even in conditions of significant undercharging. Furthermore, through simulations, Wu et al.\cite{Wu1999MonteSalts} suggest that the reduction in attraction between two surfaces at high salt concentrations originates from strong bulk ion correlations which diminish the free energy gain as the two surfaces approach each other. The main limitation of the SCL theory lies in its utilization of a mean-field PB description for the diffuse region of the double layer and the bulk. Their approach only accounts for correlations within the small condensed layer of counterions near the surface. The SCL theory's inability to capture the spatially varying ion correlation from the surface to the bulk prevents correct modeling of the structure of the overlapping double layers and their free energy. Thus, they fail to explain the underlying cause of the non-monotonic strength of like-charge attraction with respect to multivalent salt concentration. \par

Bazant et al.\cite{Bazant2011DoubleCrowding, Misra2019TheoryElectrolytes} expressed the ion correlation energy in terms of the gradient of electrostatic potential. Since the potential gradient vanishes in the bulk, this theory cannot model the bulk ion correlations correctly and thus is unable to capture the non-monotonic salt concentration dependence of like-charge attraction. Another class of methods to model ion correlations is based on the Ornstein-Zernike equation like in liquid-state integral equation theories\cite{Kjellander1988Double-LayerSwelling,Kjellander1988SurfaceElectrolytes,Rouzina1996MacroionCloud}. Recently, using such a method, Suematsu et al.\cite{Suematsu2018CiteAs} captured the non-monotonic nature of attraction. However, inhomogeneous ion-ion correlation in the interface was approximated with the bulk form of the correlation function. It is well-known that such local-density approximations sometimes give unphysical results\cite{Evans1992DensityFluids,Gillespie2011EfficientlyTheory} preventing their generalization to explain other EDL-related phenomena\cite{Weiss1998RelevanceTheory,Agrawal2022OnLayers}. In addition, solving the complete Ornstein-Zernike equation in an inhomogeneous system is numerically challenging, especially for trivalent salts\cite{Suematsu2018CiteAs}. \par 

It is also important to highlight the field-theoretic approach of Netz et al.\cite{Netz2000BeyondFunctions,Netz2001ElectrostatisticsTheory,Naji2004AttractionLimit} in modeling like-charge attraction. They used a perturbative expansion in terms of electrostatic coupling parameter $\Xi = 2\pi q^3l_\mathrm{B}^2\sigma$ to capture the coulombic energy gain due to ion correlations. Here, $q$ is the counterion valency, $l_\mathrm{B}$ is the Bjerrum length, and $\sigma$ is the surface charge density of the two surfaces. However, their analysis was limited to double layer forces in counterion-only systems, i.e., where the number of counterions in the system is fixed. These counterion-only systems are fundamentally different than salt systems, as in the latter the bulk concentration of ions also plays a critical role in determining the strength of attraction. The difference between the two kinds of systems and associated bulk correlations becomes further important as closely relevant phenomena of ion correlation-induced overcharging can only happen in salt systems. Both experiments\cite{VanDerHeyden2006ChargeCurrents,Martin-Molina2003LookingElectrolytes} and simulations\cite{Hsiao2006Salt-inducedPolyelectrolytes,Hsiao2008OverchargingStudy,Martin-Molina2008ChargeSimulations} have shown that like the strength of like-charge attraction, the strength of overcharging also depends non-monotonically on multivalent salt concentration. Therefore, it is necessary to develop a self-consistent approach to model ion correlations that can capture both overcharging and like-charge attraction in multivalent salt systems.\par 
Furthermore, in order to have a full understanding of the stability and self-assembly of charged colloids, both electrostatic and non-electrostatic effects need to be accounted for. The relative importance of the non-electrostatic effects like specific adsorption, solvation, and hydration can only be evaluated if the essential electrostatic contributions are accurately modeled. To our knowledge, existing works have not self-consistently solved the inhomogeneous electrostatic correlations from the interface to the bulk to explain like-charge attraction in multivalent salt solutions. In this article, we address this gap by leveraging the modified Gaussian renormalized fluctuation theory\cite{Wang2010FluctuationEnergy,Agrawal2022ElectrostaticElectrolytes,Agrawal2022Self-ConsistentFluids}. In our previous work, we have derived\cite{Agrawal2022ElectrostaticElectrolytes} and applied the modified Gaussian
renormalized fluctuation theory to model various ion correlation induced phenomena associated with electrical double layers. We successively captured vapor-liquid interface in ionic fluids\cite{Agrawal2022Self-ConsistentFluids}, and overcharging and charge inversion next to a single charged surface\cite{Agrawal2022OnLayers,agrawal_aiche}. Here we use the theory to study the effect of ion correlation on double layer force between two like-charged surfaces. We also explain the underlying relationship between like-charge attraction and overcharging of these two surfaces. With this approach, we aim to fully account for the effect of spatially varying ion correlations on the structure of overlapping electric double layers (EDLs) and their associated free energies. We elucidate the nature of like-charge attraction with respect to surface charge, counterion valency, salt concentration, and the addition of monovalent salts. The connection between overcharging and like-charge attraction is also examined.  \par

\section{Theory}
\label{sec:theory}

We consider a system of two similarly charged plates located at $z=0$ and $z=h$, where $h$ is the separation distance between the two plates. Both plates have a uniform surface charge density $\sigma$. The electrolyte solution with cations of valency $q_+$ and anions of valency $q_-$ is confined in the region between the plates and is connected to an outer bulk reservoir of salt concentration $c_\mathrm{b}$. The dielectric function of the system is given by $\varepsilon ({\mathbf r})$. We include the excluded volume effect of ions and solvent molecules by taking their volumes as $v_\mathrm{\pm,s} = \frac{4}{3}\pi a_\mathrm{\pm,s}^3$, where $a_\mathrm{\pm,s}$ denote the radius of ion and solvent molecules. Since the point-charge model overestimates the strength of ion correlations, ions are taken to have a finite charge spread given by distribution function $h_\pm(\mathbf{r}-\mathbf{r'})$. We apply the modified Gaussian renormalized fluctuation theory developed in our previous work\cite{Agrawal2022ElectrostaticElectrolytes, Agrawal2022Self-ConsistentFluids} to model this system. The fluctuation of the electrostatic potential beyond the mean-field level is treated using a variational approach. The consequence of this fluctuation effect is the self-energy of ions. The ion correlation is one component of the self-energy. This theory yields the following set of self-consistent equations for the electrostatic potential $\psi(z)$, ion concentration $c_\pm(z)$, self-energy of ions $u_\pm(z)$, and electrostatic correlation function $G$
 \begin{equation}
{-\nabla.[\epsilon(z)\nabla\psi(z)]} = \sigma\delta(z) + \sigma\delta(z-h) + q_{+}{c_{+}(z)} - q_{-}{c_{-}(z)} 
\label{eq:psi}
\end{equation}
\begin{eqnarray}
{c_{\pm}(z)}= \frac{ \mathrm{e}^{\mu_{\pm}}}{v_{\pm}}\exp[\mp q_{\pm}\psi(z) - u_{\pm}(z) -v_{\pm}\eta(z)]
\label{eq:conc}
\end{eqnarray}
\begin{eqnarray}
{\textit{u}_{\pm}(\mathbf{r})}=\frac{q_{\pm}^2}{2}\int d\mathbf{r}'d\mathbf{r}'^{\prime}h_{\pm}(\mathbf{r'},\mathbf{r})G(\mathbf{r'},\mathbf{r'^{\prime}})h_{\pm}(\mathbf{r'^{\prime}},\mathbf{r})
\label{eq:selfe}
\end{eqnarray} 
\begin{eqnarray}
{-\nabla_{\mathbf{r'}}.[\epsilon(\mathbf{r'})\nabla_{\mathbf{r'}}G(\mathbf{r'},\mathbf{r''})]} + 2I(\mathbf{r'})G(\mathbf{r'},\mathbf{r''}) = \delta(\mathbf{r'}-\mathbf{r''})
\label{eq:greens}
\end{eqnarray}
where $2I(\mathbf{r'})= \epsilon(\mathbf{r'})\kappa^2(\mathbf{r'}) = c_{+}(\mathbf{r'})q_{+}^2 + q_{-}^2c_{-}(\mathbf{r'})$, $\epsilon({\mathbf{r'}})=kT\varepsilon_{0}\varepsilon ({\mathbf r'})/e^2$ is the scaled permittivity with $\varepsilon_{0}$ as the vacuum permittivity and $e$ as the elementary charge. $\mu_{\pm}$ are chemical potentials of ions which are to be determined from the bulk salt concentration $c_\mathrm{b}$. $\eta(z)$ is the incompressibility field accounting for the excluded volume effect and is given by
\begin{equation}
\eta(z) =  -\frac{1}{v_s}{\ln[1 - v_+c_+(z) -v_- c_-(z)]}
\label{eq:eta}
\end{equation}
The resulting expression for the grand free energy ${W}$ is
\begin{equation}
\begin{split}
W & = \int d{z}\left[\frac{1}{2}\psi\left(\sigma\delta(z) + \sigma\delta(z-h) - q_+c_+ +q_-c_- \right) \right]
\\ & \quad + \int d{z}\left[ -c_+- c_- - \frac{(1 - c_+v_+-c_-v_-)}{v_s}\right]
\\ & \quad + \int d{z}\left[\frac{\ln(1 - c_+v_+-c_-v_-)}{v_s}\right]
\\  &\quad + \frac{1}{2}\int d{z}\int d{\bf{r'}}\int d{\bf{r'^{\prime}}}\int_{0}^{1}d\tau[G({\bf{r'}},{\bf{r'^{\prime}}},\tau)  \\ & \quad - G({\bf{r'}},{\bf{r'^{\prime}}})]\left[\sum_{i=+,-}q_i^2c_i(z) h_i({\bf{r'}}-{\bf{r}})h_i({{\bf{r'^{\prime}}}-\bf{r}})\right]
\end{split}
\label{eq:minfe}
\end{equation}
where the second and third lines account for the translational entropy of ions and solvent molecules. The fourth and fifth lines are the contribution explicitly from ion-ion correlations, with $\tau$ the charging variable. $\tau = 0$ corresponds to the case when all the ions in the system are neutral and have zero charge and $\tau = 1$ implies that all ions are fully charged to their valencies. The differential equation for $G(\mathbf{r'},\mathbf{r'^{\prime}},\tau)$ is given by
\begin{equation}
\begin{split}
{-\nabla_{\mathbf{r'}}.[\epsilon(\mathbf{r'})\nabla_{\mathbf{r'}}G(\mathbf{r'},\mathbf{r''},\tau)]} + 2\tau I(\mathbf{r'})G(\mathbf{r'},\mathbf{r''},\tau) = \delta(\mathbf{r'}-\mathbf{r''})
\end{split}
\label{eq:greens_tau}
\end{equation}
$G(\mathbf{r'},\mathbf{r'^{\prime}},\tau=0)$ represents the free space $G$ without any mobile charged particles and $G(\mathbf{r'},\mathbf{r'^{\prime}},\tau=1)$ is the correlation function when all ions are charged, as in Eq. \ref{eq:greens}. Hence the last term in Eq. \ref{eq:minfe} accounts for the electrostatic correlation contribution to free energy via the charging method where the charge is continuously added to the ions. A detailed derivation can be found in the previous work of Wang and Wang\cite{Wang2015OnSurfaces}. The disjoining pressure between the two plates can thus be calculated as: 
\begin{eqnarray}
P = -\left(\frac{\partial W}{\partial h}\right)_{\mu_\mathrm{\pm}} - P_\mathrm{b}
\label{eq:force}
\end{eqnarray}
where $P_\mathrm{b}$ is the bulk osmotic pressure accounting for the reservoir to which the two plate system is connected. \par 

It is important to understand the physical meaning behind the set of self-consistent equations in our theory. Eq. \ref{eq:psi} iis the standard differential form of Gauss' law and the net charge here includes the two surface charges and the charge from mobile cations and anions. Eq. \ref{eq:conc} is the modified Boltzmann distribution for mobile ions where the self-energy $u_\pm$ accounts for the spatially varying ion correlation and dielectric permittivity effect and $\eta$ accounts for the excluded volume effect of ions and solvent molecules. provides the relation between $u_\pm$ and the correlation function $G$. Eq. \ref{eq:greens} calculates the interaction between these two point charges at $\mathbf{r'}$ and $\mathbf{r''}$ for a given ionic environment. A decomposition scheme for \textit{G} developed in our previous work is used to solve for the spatially varying $u(\bm{r})$\cite{Agrawal2022Self-ConsistentFluids}. We decouple the short-range contribution $G_\mathrm{S}$ associated with the local electrostatic environment and the long-range contribution $G_\mathrm{L}$ associated with spatially varying ionic strength and dielectric permittivity as follows
\begin{eqnarray}
G(\mathbf{r'},\mathbf{r''}) = G_\mathrm{S}(\mathbf{r'},\mathbf{r''}) + G_\mathrm{L}(\mathbf{r'},\mathbf{r''})
\label{eq:greens_dec}
\end{eqnarray} 
The formulation is described in detail in Ref. \cite{Agrawal2022Self-ConsistentFluids} and leads to the following expression for $u_\mathrm{\pm}$
\begin{eqnarray}
\textit{u}_{\pm}(\mathbf{r})=  \frac{q_{\pm}^2}{2}\int _{\mathbf{r}',\mathbf{r}'^{\prime}}h_{\pm}G_\mathrm{S}h_{\pm} + \frac{q_{\pm}^2}{2}G_\mathrm{L}(\mathbf{r},\mathbf{r})
\label{eq:selfe_hybrid}
\end{eqnarray} 
The differential equation for $G_\mathrm{S}$ is based on Eq. \ref{eq:greens} and uses the local ionic strength $I(\mathbf{r})$ and dielectric permittivity $\epsilon(\mathbf{r})$, which yields 
\begin{equation}
-\epsilon(\mathbf{r}){\nabla_{\mathbf{r'}}^2G_\mathrm{S}(\mathbf{r'},\mathbf{r'^{\prime}})} + 2I(\mathbf{r})G_\mathrm{S}(\mathbf{r'},\mathbf{r'^{\prime}}) = \delta(\mathbf{r'}-\mathbf{r'^{\prime}})
\label{eq:greens_short}
\end{equation}
$G_\mathrm{S}$ has a Debye-H\"{u}ckel style analytical form,
\begin{eqnarray}
G_\mathrm{S}(\mathbf{r'},\mathbf{r'^{\prime}})  =  \frac{\mathrm{e}^{-\kappa(\mathbf{r})|\mathbf{r'}-\mathbf{r'^{\prime}}|}}{4\pi\epsilon(\mathbf{r})|\mathbf{r'}-\mathbf{r'^{\prime}}|}
\label{eq:gsol_short}
\end{eqnarray} 
$G_\mathrm{L}$ is obtained by subtracting Eq. \ref{eq:greens_short} from Eq. \ref{eq:greens} 
\begin{eqnarray}
{-\nabla_{\mathbf{r'}}.[\epsilon(\mathbf{r'})\nabla_{\mathbf{r'}}G_\mathrm{L}(\mathbf{r'},\mathbf{r'^{\prime}})]} + 2I(\mathbf{r'})G_\mathrm{L}(\mathbf{r'},\mathbf{r'^{\prime}})= S(\mathbf{r'},\mathbf{r'^{\prime}})
\label{eq:greens_long}
\end{eqnarray}
where the non-local source term $S$ is
\begin{equation}
\begin{split}
S(\mathbf{r'},\mathbf{r'^{\prime}})  & = \nabla_{\mathbf{r'}}.((\epsilon(\mathbf{r'})-\epsilon(\mathbf{r}))\nabla_{\mathbf{r'}}G_\mathrm{S}(\mathbf{r'},\mathbf{r'^{\prime}}))\\ & \quad - 2(I(\mathbf{r'}) - I(\mathbf{r}))G_\mathrm{S}(\mathbf{r'},\mathbf{r'^{\prime}}) 
\end{split}
\end{equation}
Using a mathematically convenient Gaussian form for $h_\pm(\mathbf{r}-\mathbf{r'})$, the short-range contribution to self energy $u_{\pm,s}$ (the first term in Eq. \ref{eq:selfe_hybrid}),  becomes\cite{Agrawal2022Self-ConsistentFluids} 
\begin{equation}
\begin{split}
u_\mathrm{\pm,S}(\mathbf{r}) & = \frac{q^2_\pm}{8\pi\epsilon(\mathbf{r})a_\mathrm{\pm}} -\frac{q^2_\pm \kappa(\mathbf{r})}{8\pi\epsilon(\mathbf{r})}\\ & \quad \times\exp\left(\frac{{a_\pm^2\kappa(\mathbf{r})}^2}{\pi}\right)\mathrm{erfc}\left(\frac{{a_\pm\kappa(\mathbf{r})}}{\sqrt{\pi}}\right)
\end{split}
\label{eq:selfe_short}
\end{equation}
Furthermore, the long-range contribution to the self-energy is not sensitive to the details of the charge distribution function $h_\pm$, hence the second term in Eq. \ref{eq:selfe_hybrid} is evaluated in the point-charge limit. The construction of this decomposition scheme is such that $G_\mathrm{L}$ only needs to be resolved at the length scale of the EDL thus making the numerical discretization tractable. Electrostatic contributions at the length scale of the ion are accounted for in $G_\mathrm{S}$ which has an analytical solution.\par
\begin{figure*}
\captionsetup[subfigure]{labelformat=empty}
    \begin{subfigure}{0.68\columnwidth}
    \includegraphics[width=\columnwidth]{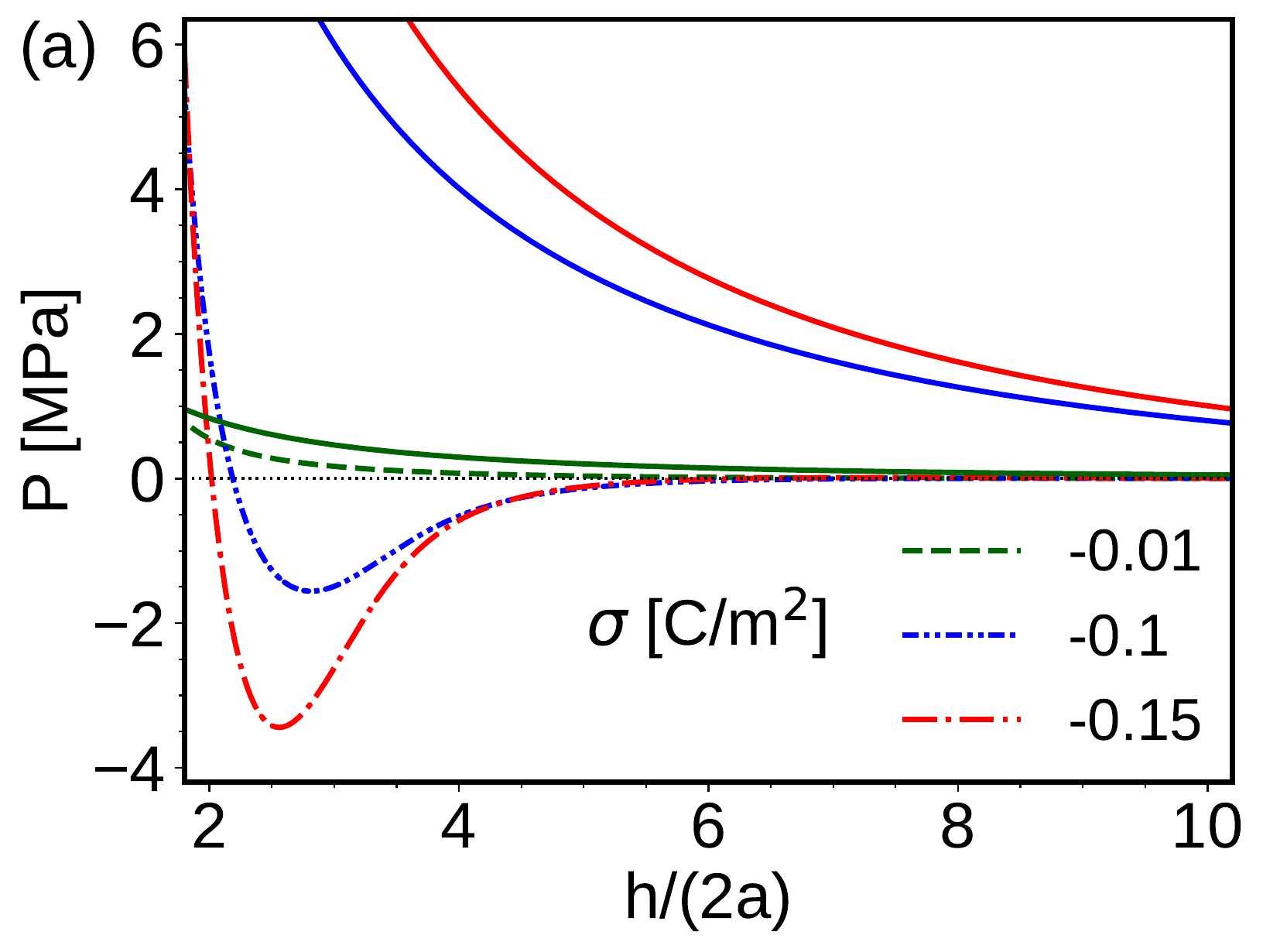}
    \caption{}
    \label{fig:force_sigma}
    \end{subfigure}  
    \begin{subfigure}{0.68\columnwidth}
        \includegraphics[width=\columnwidth]{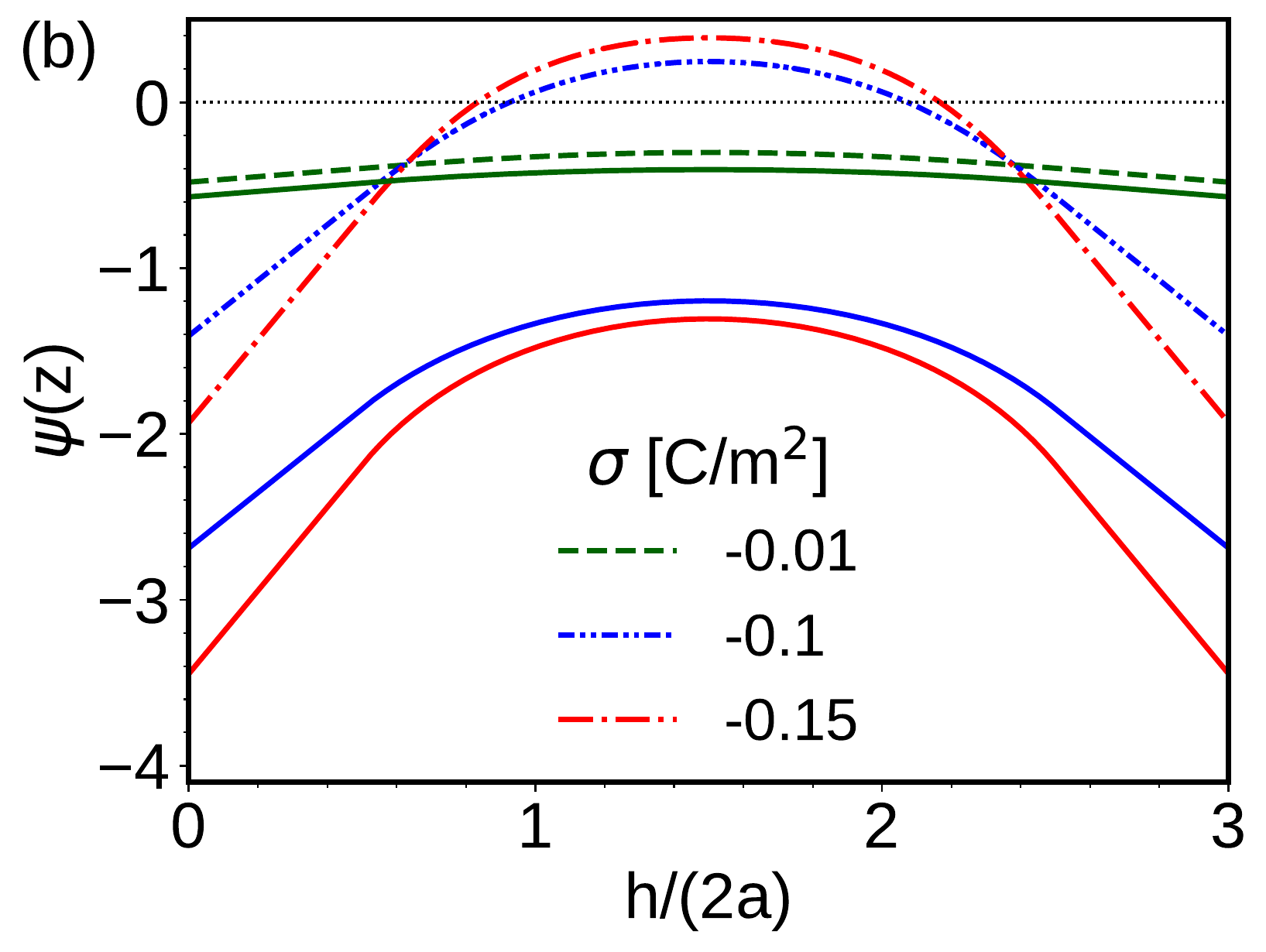}
        \caption{}
        \label{fig:psi_sigma}
    \end{subfigure} 
    \begin{subfigure}{0.68\columnwidth}
        \includegraphics[width=\columnwidth]{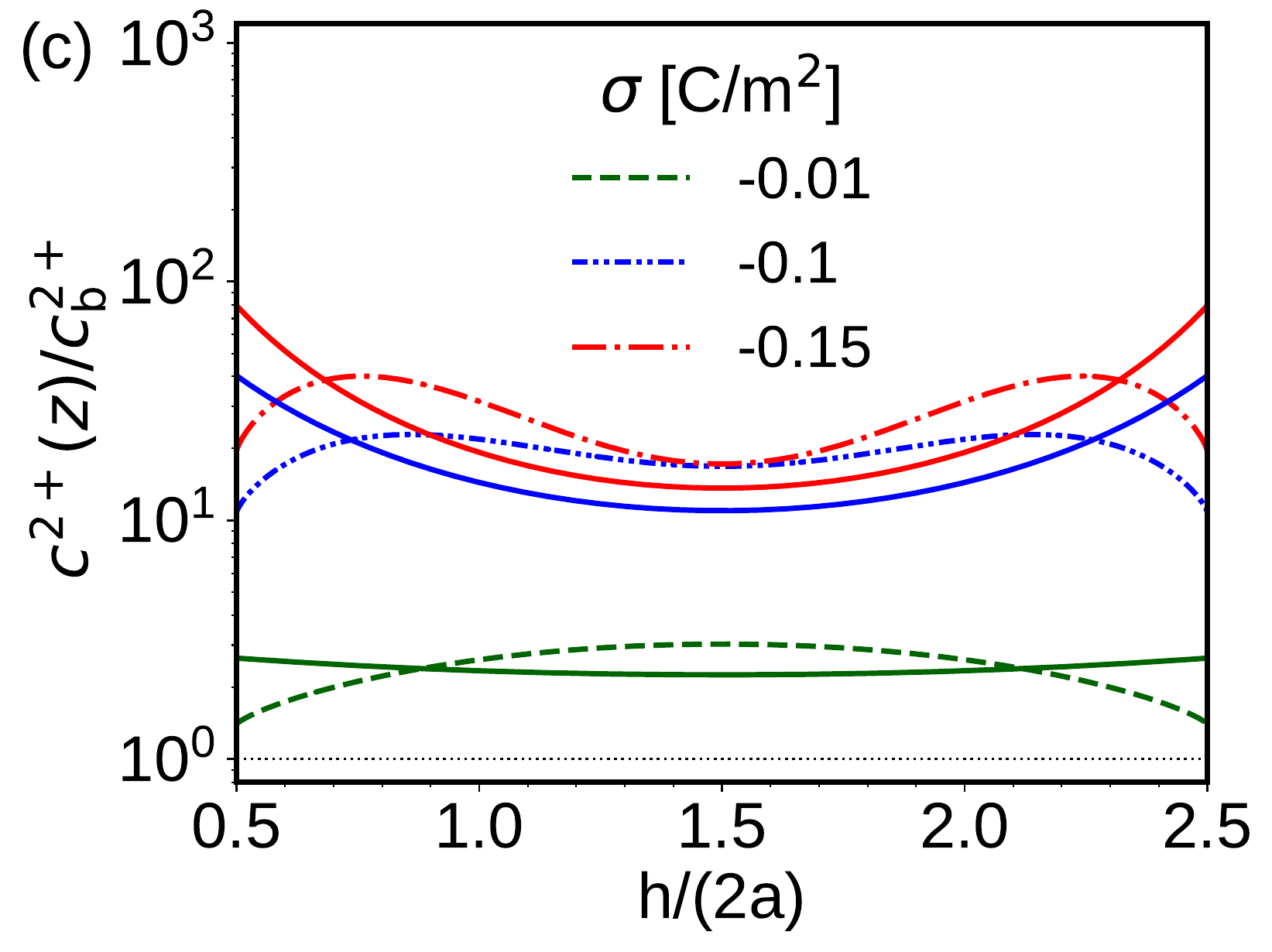}
        \caption{}
        \label{fig:conc_sigma}
    \end{subfigure} 
\caption{Continuous transition from pure repulsion to short-range attraction in overlapping double layers as surface charge density $\sigma$ increases. $q_+$ = 2, $q_-$ = 1, $c_\mathrm{b}$ = 0.1 M, $a_{\pm,s} = 1.5$ \AA. a) Disjoining pressure as a function of separation distance $h$ between the two plates. b), and c) plot electrostatic potential profiles $\psi(z)$, and counterion concentration profiles, respectively for $h = 6a$. Dashed lines represent predictions from our theory and solid lines represent mean-field Poisson-Boltzmann results.}
\label{fig:panel}
\end{figure*}
It is also important to highlight here the difference between the modified Gaussian renormalized fluctuation theory and other variational theories formulated by Netz and Orland\cite{Netz2003VariationalSystems}, and Buyukdagli and co-workers\cite{BuyukdagliVariationalNanopores,Buyukdagli2017Like-chargeMolecules}. There are three key differences between our work and other approaches. First, instead of a point-charge model for mobile ions, we consider a finite spread of ionic charge given by distribution function $h$. This finite spread avoids the overestimation of ion-ion correlation aroused by the point-charge model and also removes any divergence issues from the self-energy term. Thus this finite spread leads to the ``renormalization" of the fluctuation contribution to the free energy. The word ``fluctuation" here represents the fluctuations in the electrostatic potential caused by the constant motion of the mobile ions as a result of coulombic and thermal forces. Secondly, to deal with the dual-length scale problem associated with the calculation of self energy of an ion with finite charge spread, the ion correlation is decoupled into a short-range contribution associated with the local electrostatic environment and a long-range contribution accounting for the spatially varying ionic strength and dielectric permittivity\cite{Agrawal2022Self-ConsistentFluids}. Finally, we also account for the excluded volume effect of ions and solvent molecules through the incompressibility factor $\eta$. The second and third differences mentioned above correspond to the term ``modified" in the name of our theory, compared to the original version of the Gaussian Renormalized Fluctuation theory developed by Z.-G. Wang \cite{Wang2010FluctuationEnergy}.
In this theory, the instantaneous electrostatic potential, whose mean is $\psi$, was assumed to have a Gaussian distribution. This ``Gaussian ansatz", serves as a simplifying assumption to facilitate the calculation of the partition function and hence the free energy.

\section{Results and Discussion}
\label{sec:Results}

\begin{figure*}
    \captionsetup[subfigure]{labelformat=empty}
    \begin{subfigure}{\columnwidth}
    \includegraphics[width=\columnwidth]{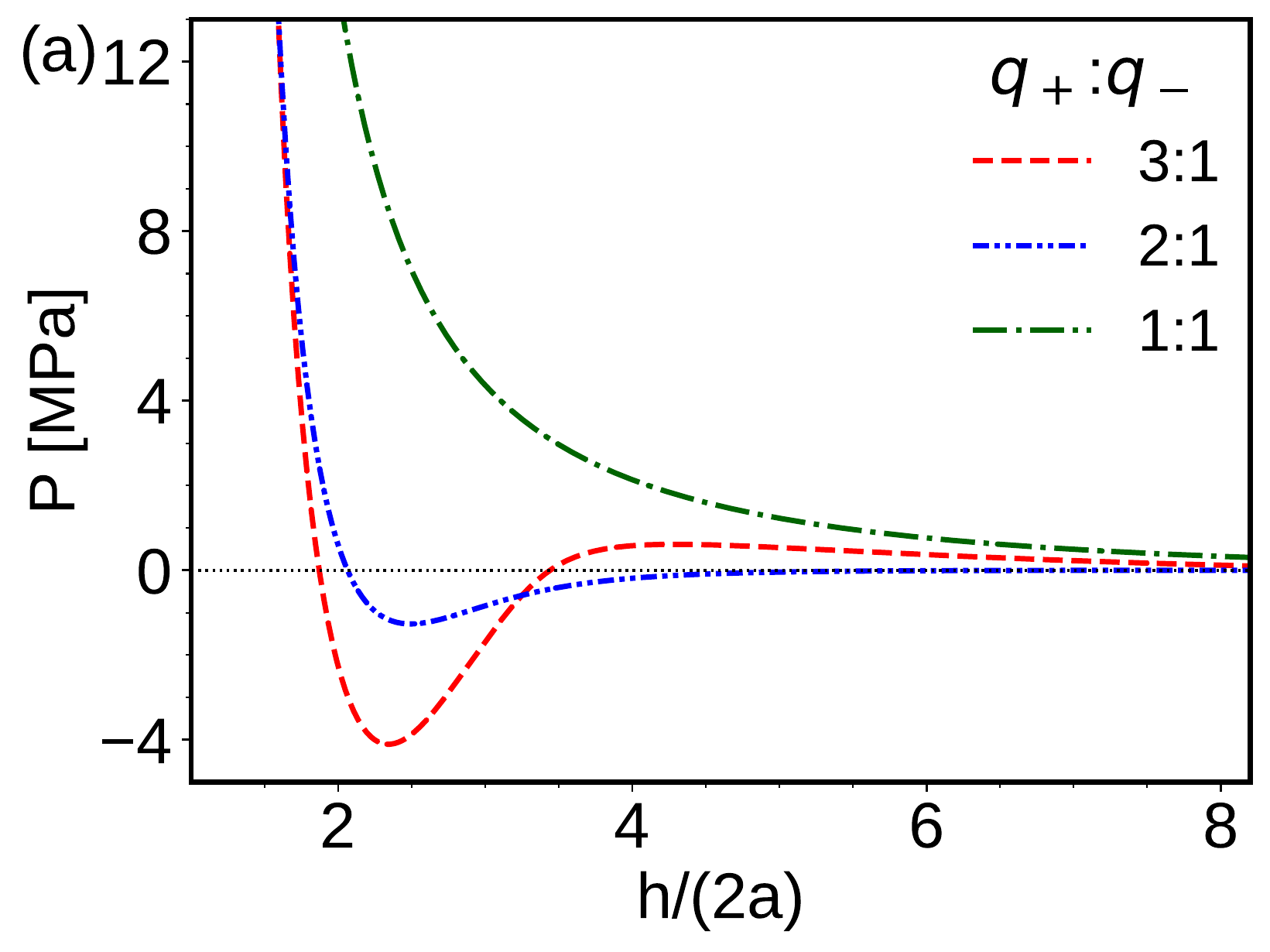}
    \caption{}
    \label{fig:force_val}
    \end{subfigure}  
    \begin{subfigure}{\columnwidth}
        \includegraphics[width=\columnwidth]{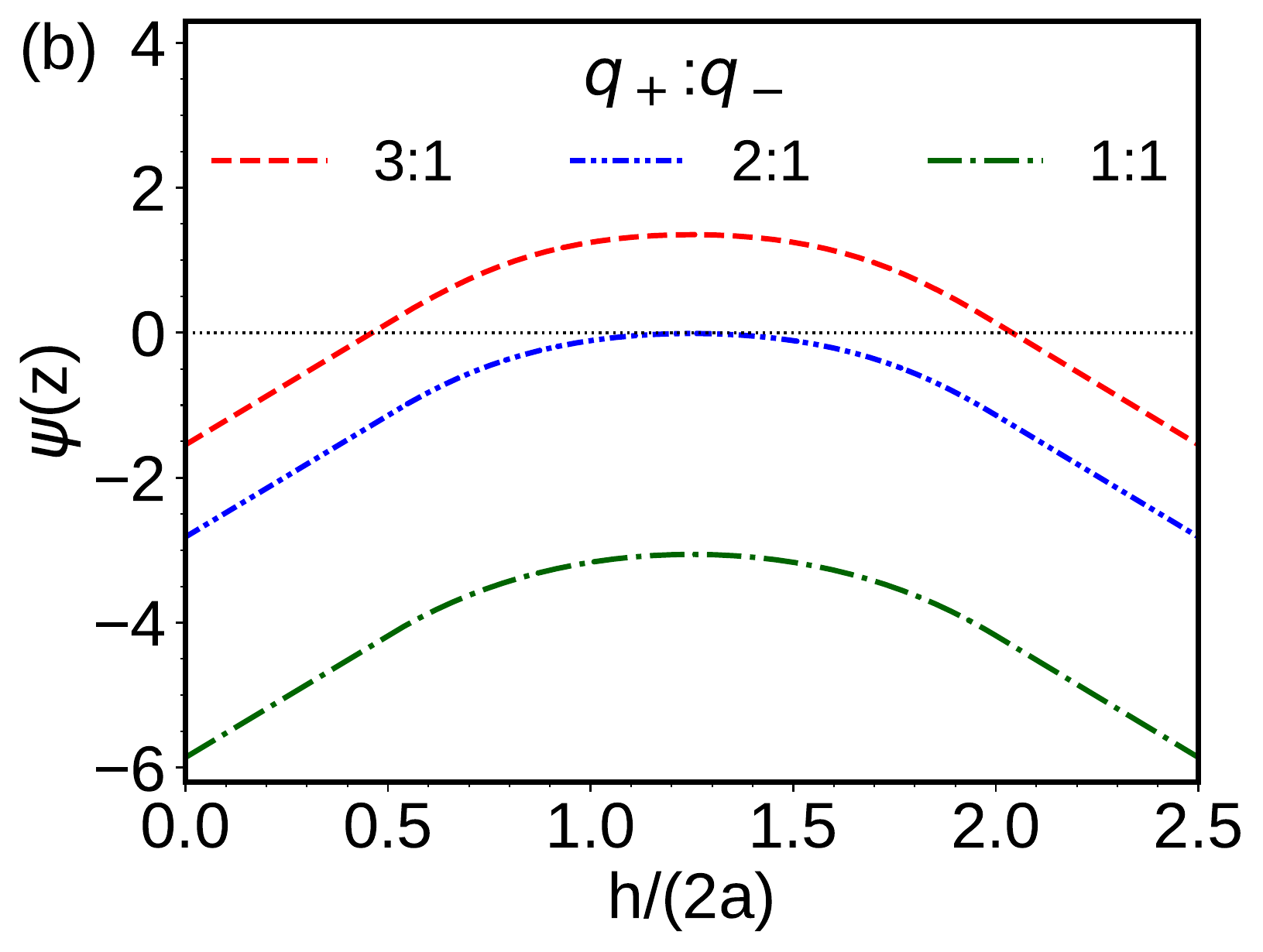}
        \caption{}
        \label{fig:psi_val}
    \end{subfigure} 
\caption{a) The effect of counterion valency on disjoining pressure \textit{P}. b) Electrostatic potential $\psi(z)$ for $h$ = $5a$. $c_\mathrm{b} = 0.1$ M, $\sigma = - 0.15$ C/m$^2$, $a_\mathrm{\pm,s} = 2.0$ \AA.}
\label{fig:val_panel}
\end{figure*}

In the current work, we study the effect of surface charge, counterion valency, and salt concentration on double layer forces. For simplicity, the dielectric function is assumed to be constant throughout the system, $\varepsilon(\bm{r}) = 78.5$, i.e., the primitive model for electrolytes. Ions and solvent molecules are considered to have the same radius $a$, which is also the minimum possible separation between an ion's center and the charged surface. In agreement with simulations\cite{Guldbrand1984ElectricalStudy, Pellenq1997ElectrostaticSimulation, IseLikeInteractionw}, our theory predicts a continuous transition from pure repulsion to like-charge attraction as surface charge density increases. Figure \ref{fig:force_sigma} plots the disjoining pressure $P$ as a function of separation distance $h$ for the case of 0.1 M divalent salt solution, $q_+$ = 2 and $q_-$ = 1. At a low $\sigma$ value of $-0.01$ C/m$^2$, the pressure remains positive (repulsion) at all separations, in quantitative agreement with the predictions of mean-field PB. Increasing $\sigma$ to $-0.1$ C/m$^2$ leads to a weak short-range attraction. This attraction gets more pronounced as $\sigma$ becomes more negative. In stark contrast, PB predicts a stronger repulsion as the magnitude of $\sigma$ increases. For all $\sigma$ values, attraction is short-ranged and occurs at a separation distance of a few ion diameters, in agreement with simulations\cite{linselca1999,Linse2000,Wu1998InteractionSolutions}. At large $h$, the force has a weak repulsive tail compared to PB, which has also been observed in  simulations\cite{Wu1998InteractionSolutions, Wu1999MonteSalts, Angelescu2003MonteAdded}. \par 

To understand the origin of like-charge attraction, electrostatic potential profiles $\psi(z)$ and counterion concentration profiles $c_\mathrm{b}^\mathrm{2+}(z)$ are plotted in Figure \ref{fig:psi_sigma} and \ref{fig:conc_sigma} respectively. The inclusion of ion-ion correlations leads to an enhanced screening of the surface charges hence reducing surface potential and the repulsive component of the free energy (first line in Eq. \ref{eq:minfe}). The remaining attractive contribution directly comes from the gain in ion-ion correlation energy described by the last term in Eq. \ref{eq:minfe}. At low surface charges, counterion concentrations are not high enough to make correlation effects important. Hence our pressure, $\psi(z)$, and $c_\mathrm{b}^\mathrm{2+}(z)$ predictions are very close to PB. It is at high surface charges where the correlation contribution to free energy starts dominating; our theoretical predictions start deviating significantly from PB and an attractive force appears. It is also worth noting that in Figure \ref{fig:conc_sigma} ion concentrations are significantly larger than the bulk at the mid-plane ($z$=$h/2$). Thus, if the pressure is calculated using the mean-field Maxwell stress, one would only get a repulsive force. The reason our theory correctly predicts the short-range attraction is the systematic inclusion of the ion correlation contribution to the free energy. Figure \ref{fig:panel} also demonstrates the lack of any causal relationship between overcharging and like-charge attraction. The curvature of $\psi(z)$ remains negative for all chosen $\sigma$ values, implying the excess of positive counterions and the absence of any overcharging of EDL. However short-range attraction is still observed at high $\sigma$. Our results elucidate that while both the phenomena of overcharging and like-charge attraction arise from ion correlations, the presence of one phenomenon does not inherently imply the presence of the other. Similarly, the existence of one is not a prerequisite for the occurrence of the other. This also confirms the simulation results of Allahyarov et al.\cite{lowenreentrant} and P.Y. Hsiao\cite{Hsiao2008OverchargingStudy}. Our modified Gaussian renormalized fluctuation theory is the first to coherently capture these two phenomena in a unified framework\cite{Agrawal2022OnLayers}. \par 

Counterion valency is another key factor in determining the strength of ion correlations and hence the double-layer force. Figure \ref{fig:force_val} plots the pressure profiles for different counterion valencies. For monovalent counterions, correlations are very weak and the force is repulsive at all length scales. In fact, for all values of $\sigma$, our theory does not predict any attraction in an aqueous monovalent salt solution at room temperature, which is consistent with observations in simulations. \cite{Wu1998InteractionSolutions, Wu1999MonteSalts,linselca1999, Linse2000, Angelescu2003MonteAdded, Zhang2016PotentialEffect}. For divalent counterions, correlation strength increases, leading to a small attractive force. For trivalent counterions, a very strong attractive force is predicted. This increasing strength of attraction is in agreement with simulations of Linse and Lobaskin \cite{linselca1999, Linse2000}, where the size of colloidal aggregates was found to be significantly larger in the presence of trivalent ions compared to divalent ions. Furthermore, $\psi(z)$ profiles in Figure \ref{fig:psi_val} show that an increase in the strength of correlations also enhances the possibility of overcharging. The EDL is not overcharged for any of the three counterions; however, both trivalent and divalent ions show like-charge attraction. This again proves the lack of any causal relationship between overcharging and like-charge attraction. \par
\begin{figure*}
\captionsetup[subfigure]{labelformat=empty}
    \begin{subfigure}{\columnwidth}
    \includegraphics[width=\columnwidth]{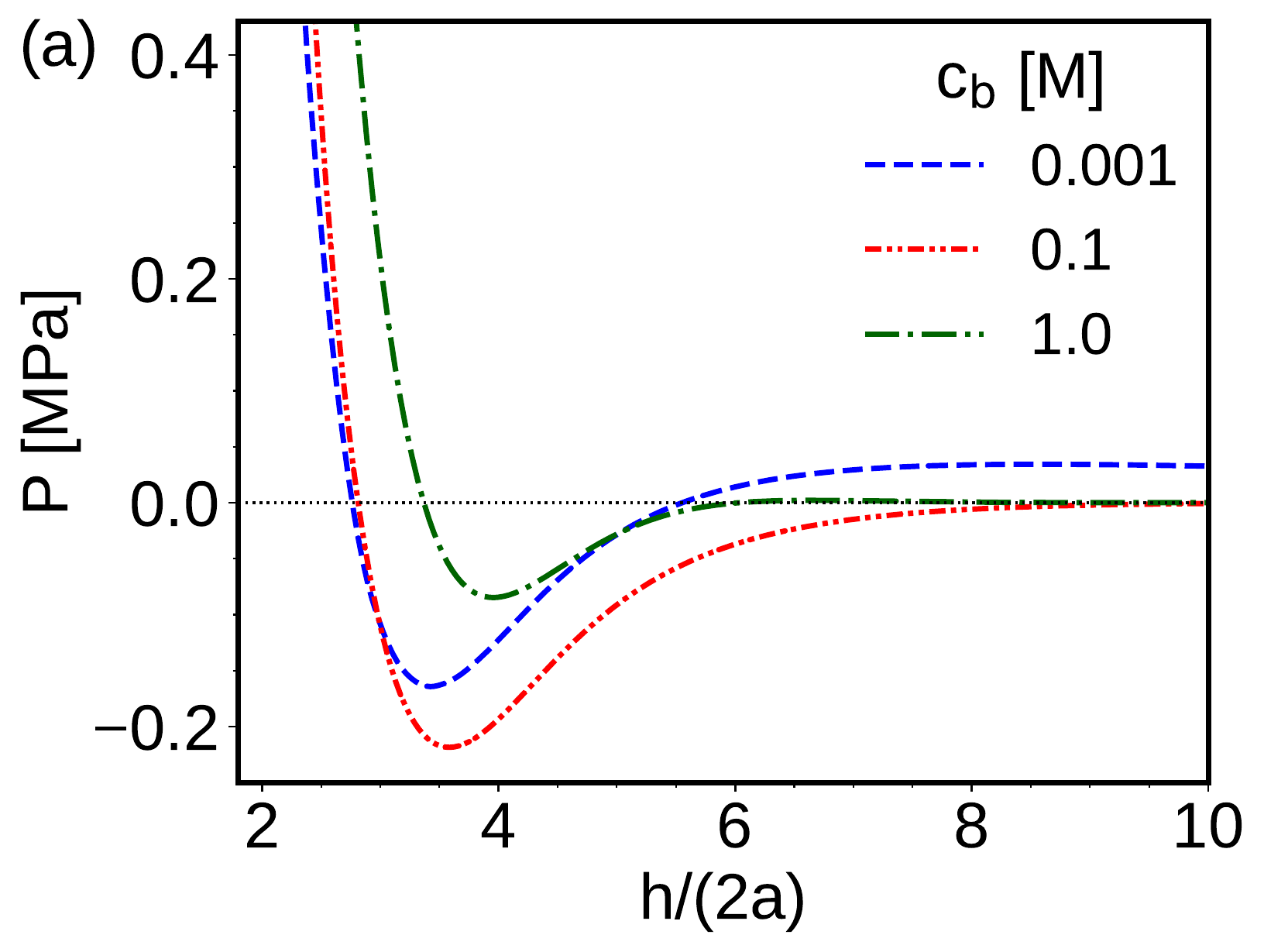}
    \caption{}
    \label{fig:force_conc}
    \end{subfigure}  
    \begin{subfigure}{\columnwidth}
        \includegraphics[width=\columnwidth]{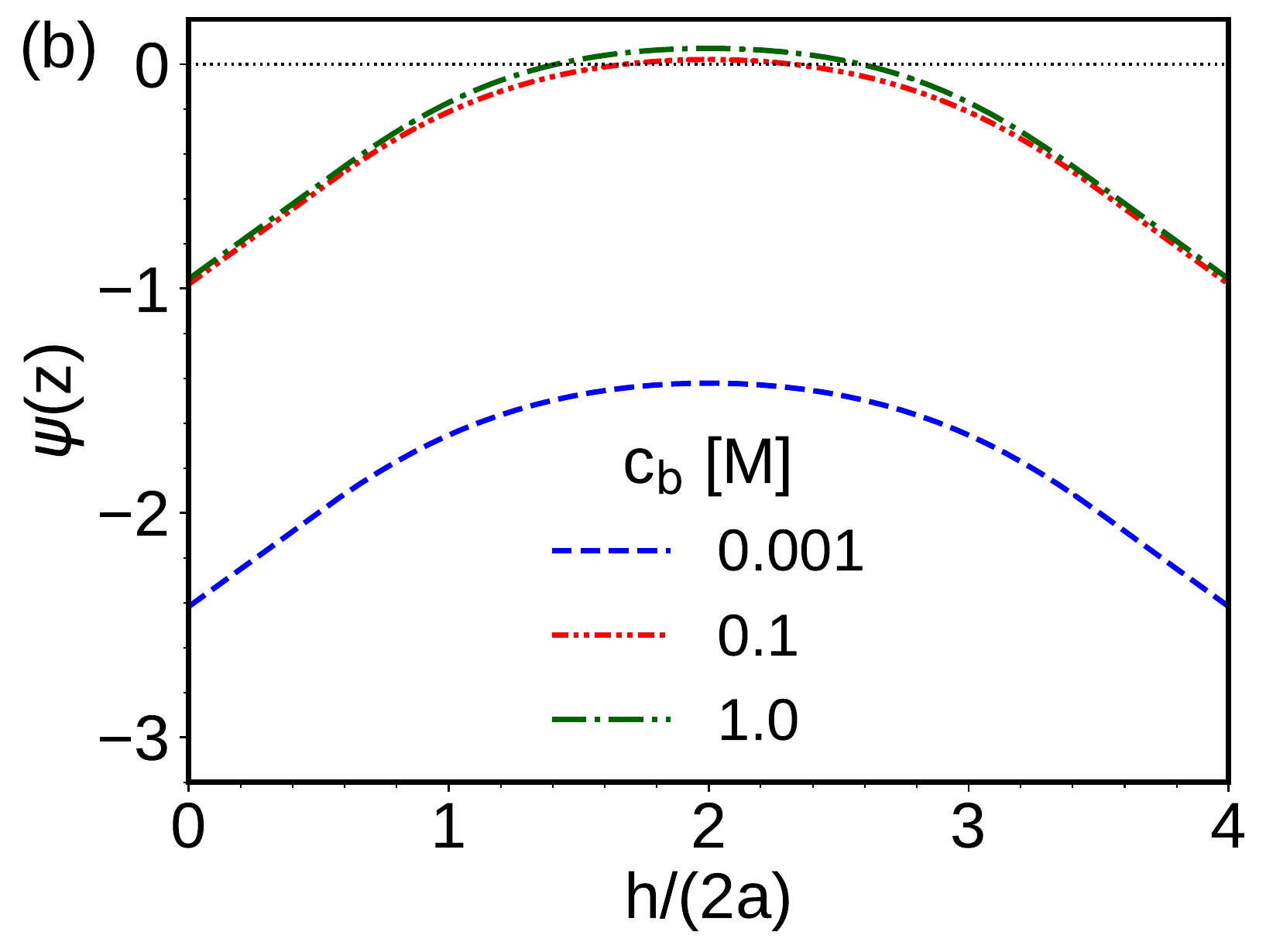}
        \caption{}
        \label{fig:psi_conc}
    \end{subfigure} 
\caption{a) Non-monotonic salt concentration dependence of the attractive force. b) Electrostatic potential profiles $\psi(z)$ at h = 8$a$, where the force is attractive for all three c$_\mathrm{b}$. $q_+$ = 2, $q_-$ = 1, $\sigma = - 0.05$ C/m$^2$, $a_\mathrm{\pm,s} = 1.5$ \AA.}
\label{fig:conc_panel}
\end{figure*}
Monte Carlo simulations have shown that the attractive force does not monotonically increase with salt concentration\cite{Wu1999MonteSalts, Angelescu2003MonteAdded}. This non-monotonic salt concentration dependence of like-charge attraction is also correctly captured by our theory. Figure \ref{fig:force_conc} shows pressure profiles for three different salt concentrations. Increasing $c_\mathrm{b}$ from 0.001 M to 0.1 M weakens the long-range repulsion because of the enhanced screening of surface charges. In addition, the attraction well deepens, as higher bulk ion concentration reduces entropy loss for ions to move to the surface. This drives more counterions to the surface leading to larger gains in the correlation energy. However, attraction is significantly reduced for $c_\mathrm{b}$ = 1.0 M . At very high salt concentrations, the strength of correlations in the bulk itself is very strong and there is no gain in correlation energy for the ions when they come from bulk to the surface. The accumulation of ions is further bounded by the excluded volumes of ions and solvent molecules. In Figure \ref{fig:psi_conc}, we also plot $\psi(z)$ for the three $c_\mathrm{b}$ at $h = 8a$, where all three concentrations show attractive force. The curvature of $\psi(z)$ remains negative for all $c_\mathrm{b}$ values implying no overcharging of the surfaces. These results are in agreement with the simulations by Pai-Yi Hsiao \cite{Hsiao2008OverchargingStudy} which showed that the redissolution of polyelectrolyte condensates at high salt concentrations can occur without any inversion of their net charge. In addition, the same simulations also showed that like-charge attraction can happen even in conditions of significant undercharging, a feature that is also captured by our theory, see the curves of $c_\mathrm{b}$ = 0.001 M in Figure \ref{fig:conc_panel}. Therefore, this non-monotonic behavior of the attractive force should be understood as a result of the competition between ion correlations and the translational entropy of the ions. \par 

Based on the Strongly Correlated Liquid theory (SCL), Shklovskii and others\cite{Nguyen2000ReentrantCounterions,Grosberg2002Colloquium:Systems,lobaskinqamhieh} argued that like-charge attraction is instead a consequence of competition between the repulsion between the net charge of the two surfaces and the attractive ion-ion correlation contribution to free energy. According to SCL theory, attraction disappears at conditions of significant undercharging and overcharging. In their framework, the net charge of the surface is calculated considering an isolated macroion surrounded by salt solution. An attempt to establish a connection between the effective charge of a macroion in isolation to the aggregation of a group of macroions in a salt solution was made. However, during the aggregation process, as the two macroions approach each other, the interactions between the two overlapping double layers will also change the effective surface charge of the two macroions. This effective charge will be different from the effective surface charge of an isolated macroion in a multivalent salt solution. Hence, we disagree with SCL theory's argument that the redissolution of aggregates at high salt concentration is an outcome of repulsion between net inverted charges of the two surfaces, particularly, as this net inverted charge was calculated for an isolated macroion. In fact, for two macroions separated by a few ion diameter separations, the two double layers completely overlap each other, and it's impossible to define the net charge of one macroion. The true reason for like-charge attraction and its non-monotonic strength with respect to salt concentration is the combination of translational entropy and the ion-ion correlation contribution to the free energy as explained in the previous paragraph. \par
\begin{figure}
\includegraphics[width=\columnwidth]{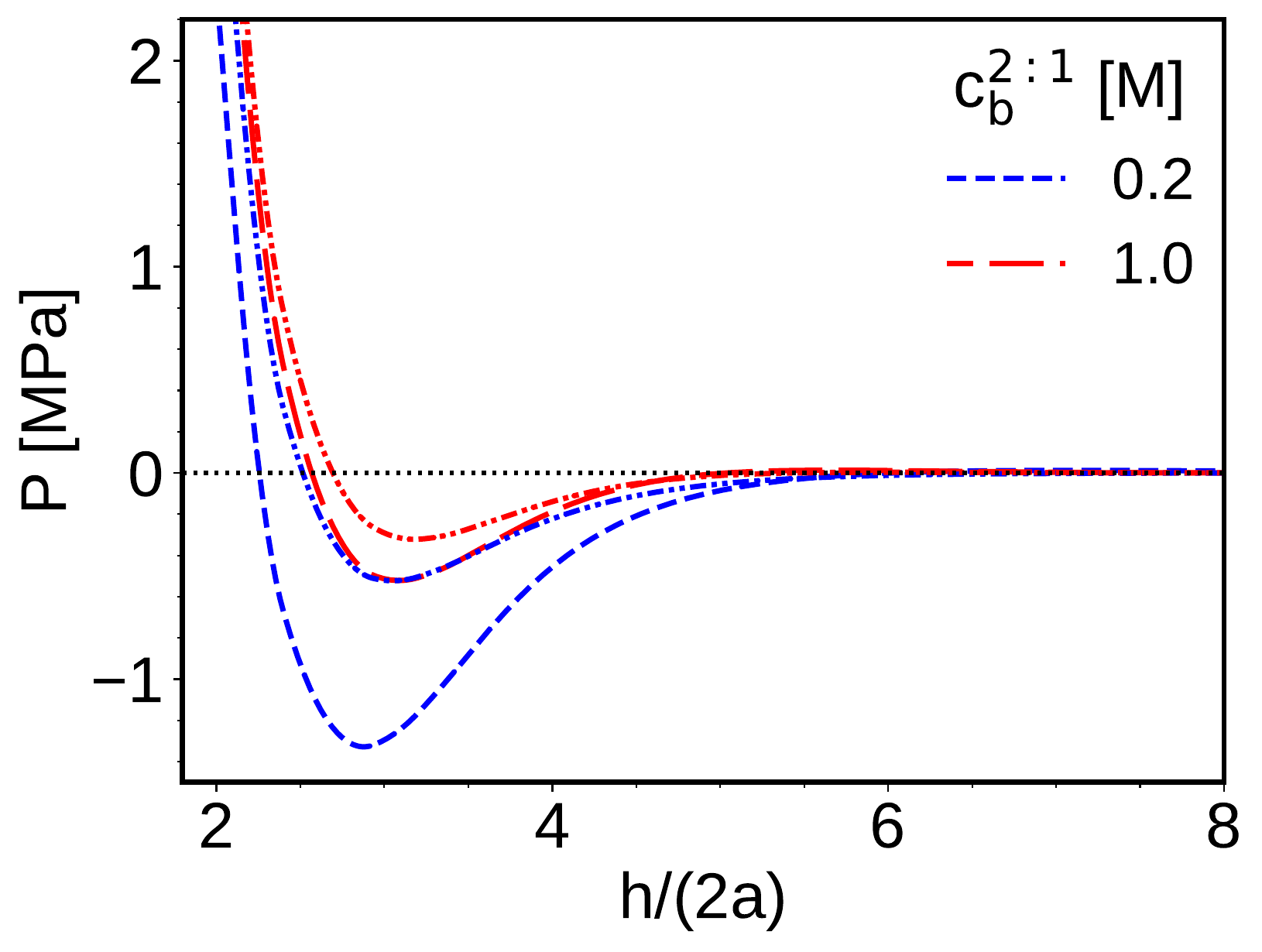}
  \caption{The effect of adding 0.5 M monovalent salt to a solution with fixed concentrations of divalent salt $c_\mathrm{b}^\mathrm{2:1}$ on like-charge attraction. The solid lines represent pressure curves for the pure divalent salt solution and the dashed lines represent pressures after monovalent salt is added. $\sigma = - 0.1$ C/m$^2$, $a_\mathrm{\pm,s} = 1.5$ \AA.}
  \label{fig:force_mix}
\end{figure}
The addition of monovalent salt to a multivalent salt solution also increases the strength of ion correlations in bulk and leads to a decrease in the strength of attraction. In Figure \ref{fig:force_mix}, 0.5 M of monovalent salt is added to 0.2 M and 1.0 M of a fixed concentration of divalent salts ($c^\mathrm{2:1}_\mathrm{b}$). For low $c^\mathrm{2:1}_\mathrm{b}$, the addition of monovalent salt leads to a significant reduction in attraction. However, as $c^\mathrm{2:1}_\mathrm{b}$ increases, the effect of monovalent salt on this suppression of attraction reduces. This effect is almost negligible for 1.0 M. At 1.0 M of $c^\mathrm{2:1}_\mathrm{b}$, the strength of bulk correlations due to divalent salt is high enough that further addition of monovalent salt does not lead to any change in attraction. The curve for the salt mixture completely overlaps with the case of pure divalent salt. A similar dependence of adding monovalent salt has been indicated in the solubility measurements of charged colloids \cite{Pelta1996DNACobalthexamine, Raspaud1998PrecipitationBehaviour, OlveraDeLaCruz1995PrecipitationSalts}. Experiments found that the addition of monovalent salt increases the critical multivalent salt concentration to initiate aggregation. However, 
the re-dissolution of charged aggregated at high multivalent salt concentrations was shown to be insensitive to monovalent salt. \par

\section{Conclusion}

We apply the modified Gaussian Renormalized Fluctuation theory to study the phenomenon of ion correlation-driven like-charge attraction. The theory self-consistently captures the effects of inhomogeneous ion correlations on the structure of overlapping double layers and their free energy. The systematic inclusion of correlation in the free energy is key to successfully modeling like-charge attraction. We predict a continuous transition from pure repulsion to short-range attraction as the surface charge density increases. At high surface charges, more ions migrate to the surface, leading to a gain in correlation energy. This attraction is found to be absent for monovalent salts and becomes more pronounced as the counterion valency increases. We demonstrate that overcharging is not a necessary condition for like-charge attraction. Like-charge attraction is observed for both overcharged and normal double layers. Furthermore, our theory is also able to capture the non-monotonic salt concentration dependence of the attractive force as a consequence of the competition between ion correlations and translational entropy. The reduction of attraction at high salt concentrations could be a contributing factor to the phenomena of reentrant condensation. Our theoretical predictions are in agreement with the simulation and experimental results. The modified Poisson-Boltzmann structure of equations in our theory makes it convenient to include ion-specific van der Waals and soft repulsive interactions in the prediction of double layer structure and free energies, as has been recently shown in Seal et al.\cite{gupta_vanderwaals}. Finally, the self-consistent quantification of the essential electrostatic contributions presented here paves the way toward a complete understanding of interfacial forces. 



\begin{acknowledgments}
Acknowledgment is made to the donors of the American Chemical Society Petroleum Research Fund for partial support of this research. The authors thank Dr. Dimitrios Fraggedakis of UC Berkeley for his insightful comments on the results. This research used the computational resources provided by the Kenneth S. Pitzer Center for Theoretical Chemistry at UC Berkeley.

\end{acknowledgments}

\section*{Data Availability Statement}

The data that support the findings of this study are available from the corresponding author upon reasonable request.

\bibliography{lca_jcp}

\end{document}